\documentclass[global,referee]{svjour}

\usepackage{soul}
\usepackage{color}
\usepackage{graphicx}
\usepackage[draft]{hyperref}
\usepackage{amsmath}
\usepackage{amssymb}
\usepackage{cite}
\begin{document}


\title{Proton irradiation robustness of dielectric mirrors for high-finesse Fabry-P\'erot resonators in the near-infrared spectral range}


\author{Qun-Feng Chen\inst{1} \and Alexander Nevsky\inst{1} \and Stephan Schiller\inst{1} \and Erwin Portuondo Campa\inst{2} \and Steve Lecomte\inst{2} \and David Parker\inst{3}}

\institute{Institut f\"{u}r Experimentalphysik, Heinrich-Heine-Universit\"{a}t D\"{u}sseldorf, 40225 D\"{u}sseldorf, Germany
\and
Centre Suisse d'Electronique et Microtechnique SA, 2002 Neuch\^{a}tel, Switzerland
\and
School of Physics and Astronomy, University of Birmingham, B15 2TT Birmingham, United Kingdom
}
\date{}
\maketitle
\begin{abstract}
We demonstrate that a proton irradiation with fluences of $3.6\times10^{10}$/cm$^{2}$ at low energy ($<$ 36 MeV) and $1.46 \times 10^{10}$/cm$^{2}$ at high energy (40 MeV and 90 MeV combined) on the dielectric mirrors of Fabry-P\'erot cavities with a finesse of about 700 000 causes less than 5\% change in the finesse. Furthermore, no influence on the coupling efficiency to the cavities was observed, the efficiency being approximately 70\%. The irradiation was carried out with a spectrum approximating the proton energy spectrum of a highly elliptic Earth orbit with duration of 5 years, proposed for the Space-Time Explorer and Quantum Equivalence Space Test (STE-QUEST) mission [\url{http://sci.esa.int/ste-quest/}].
\end{abstract}


\section{Introduction}
 \noindent 
Precision experiments in space are a powerful approach for testing fundamental notions of space and time. It has been proposed to operate atomic clocks, including optical clocks, on Earth-bound and deep-space orbits for testing the gravitational time dilation effect and the Shapiro effect predicted by General Relativity \cite{Schiller:2009, Wolf:2009, STEQUEST, Schiller:2012, Ashby:2009}. In such missions, ultra-stable lasers are foreseen to be used as the local oscillators to probe the clock transition of the atoms. In the case of an optical transition, the probing is performed by the laser directly. To interrogate a microwave clock transition, the laser radiation with ultrastable frequency can be converted to a spectrally very pure microwave signal, using a femtosecond frequency comb \cite{Lipphardt:2009}. Since the short-term frequency stability of an ultra-stable laser can be over 100 times better than that of the currently best quartz oscillator, microwave radiation with an extremely low level of phase noise can be generated.

Laser radiation with ultra-stable frequency is generated by stabilizing a laser to a high-finesse optical resonator {located in a high-vacuum chamber}. {In the laboratory, the resonator properties (e.g. linewidth, throughput, and frequency stability etc.) are very stable and do not degrade, even over years of operation, while in space, because of strong particle (electrons and protons) radiation, the resonator properties } may degrade in time, and therefore the stability of the laser may also degrade. Several previous works show the influence of particle irradiation on the dimensional and/or elastic properties of glass, ceramics, and multi-layer EUV mirrors \cite{Higby:1988, Zhu:1994, Peng:2012}. In a previous study, performed with gamma rays, no influence of irradiation on a cavity with finesse of 1500 at 1 $\mu$m wavelength was observed \cite{Heine:2007}. To our knowledge, no result on the influence of proton irradiation on high-finesse cavities has been reported until now. Proton radiation of the typical energies encountered in the radiation belt is difficult to attenuate strongly by shields of practical thickness, and so this radiation represents a potential risk for components used in space. 

In this paper, we present the results of a test of the influence of proton irradiation on high-finesse infrared Fabry-Perot cavity mirrors. We have paid special attention towards providing a proton irradiation energy spectrum ``engineered'' according to the differential proton fluence of the Space-Time Explorer and Quantum Equivalence Space Test (STE-QUEST) mission \cite{STEQUEST}. In this mission, the satellite will operate in a highly elliptic Earth orbit for a duration of up to 5 years. We test the influence of the high-energy part of the spectrum for an equivalent duration of 1, 3, and 5 years, and of the lower-energy part of the spectrum for an equivalent duration of 5, 10, and 15 years. The latter two durations were implemented in order to increase the likelihood of a measurable effect.

\section{The proton radiation spectrum in space}
The STE-QUEST orbit is highly elliptical, with a semi-major axis of 32 000 km. The orbit period is 16 h and the orbit crosses the van Allen radiation belt. Details of the orbit can be found in \cite{STEQUESTENV}. Along the orbit, radiation both from trapped particles and from solar particles is encountered. In the framework of a preliminary study of this orbit \cite{STEQUESTENV}, the trapped proton radiation was calculated by using the standard model AP-8 \cite{Sawyer:1976}. The solar proton spectrum was calculated by the ESP model \cite{Xapsos:2000}. In this work, we use the simulation data given in \cite{STEQUESTENV} and consider the sum of the trapped proton and the solar proton radiation spectra and denote it by ``total'' radiation.

\begin{figure}[htb]
	\centering
	\includegraphics[height=4cm]{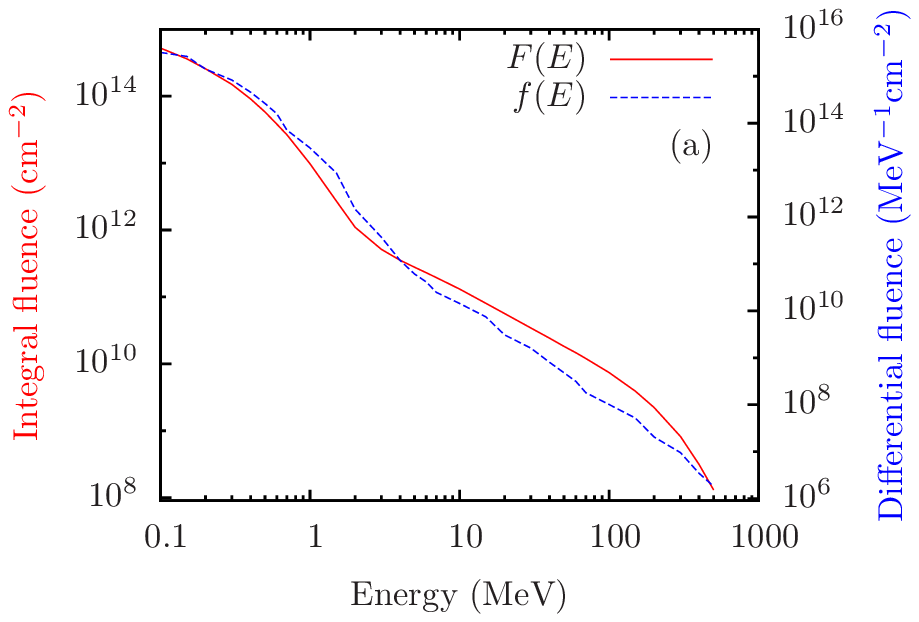}~~~\includegraphics[height=4cm]{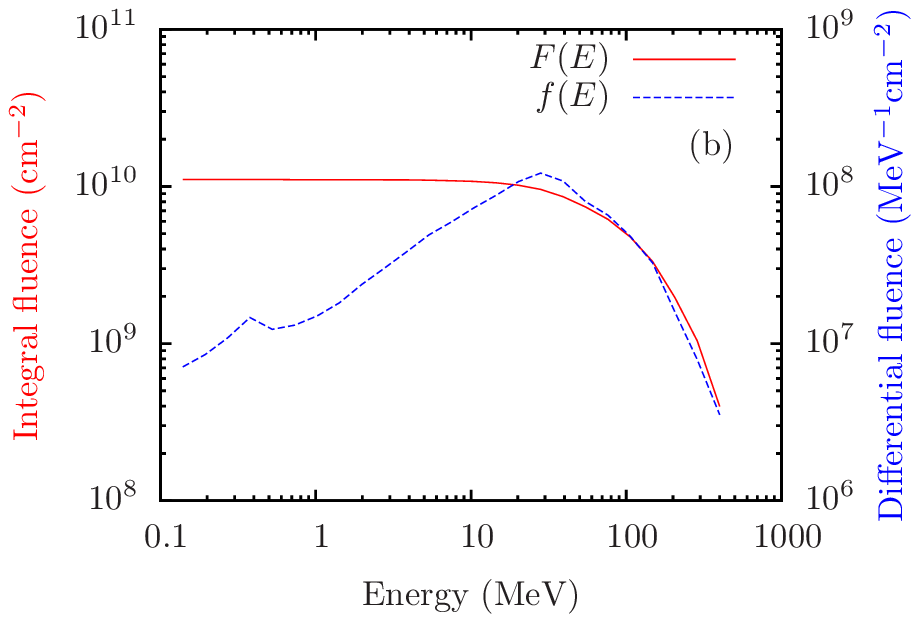}
	\caption{(a) The spectrum of the total proton radiation in the 5 year-long STE-QUEST mission. (b) The corresponding proton radiation spectrum after 10 mm aluminum shielding. }
	\label{fig:orbitFl}
\end{figure}

Figure \ref{fig:orbitFl} (a) shows the integral and differential fluence of the total protons integrated over 5 years and along the orbit. This integral fluence $F(E)$ is related to the differential fluence $f(E)$, the impinging particle number per unit area and energy interval, by
\begin{equation*}
	F(E)=\int_{E}^{+\infty}f(E')\;dE'\,.
\end{equation*}

In the STE-QUEST payload scenario, the optical cavity will be located inside a vacuum chamber, naturally providing partial shielding. Moreover, typically the vacuum chamber will be placed inside the satellite and therefore the satellite structure as well as other equipment will provide additional shielding. Therefore, it is conservative to assume that there exists a shielding equivalent to 10 mm aluminum. The shielding effect of 10 mm aluminum is calculated by using the software MULASSIS \cite{MULASSIS}. The spectrum of the proton fluence after the shielding is shown in Fig. \ref{fig:orbitFl} (b).

\section{The mirrors}

The mirrors investigated in this work are high-reflectivity mirrors produced by Advanced Thin Films (Boulder, USA). The substrate material is fused silica, the diameters are 0.5 inch, the radii of curvature are 0.5 m, and they are coated for high reflection at 1.5 $\mu$m. The coatings of the mirrors consist of 41 layers of Ta$_{2}$O$_{5}$ and SiO$_{2}$. The wavelength 1.5 $\mu$m is a candidate wavelength for a laser system suitable to provide an ultra-low-noise and ultrastable microwave, in particular because there exist space-qualified lasers at 1.5 $\mu$m and robust fiber-based erbium frequency combs at the same wavelength. 

\section{The irradiation runs}

The irradiation was performed in two steps, providing the high-energy part and the low-energy part of the spectrum, respectively. The first proton irradiation test was carried out at the Paul-Scherrer-Institut (PSI) in Villigen, Switzerland, the second at the University of Birmingham, United Kingdom. The time interval between the two irradiations was 5 months. We note that during a space mission, the irradiation would have a very different time dependence as compared to our combined accelerator irradiations. Since we measured the effect of each partial irradiation within only a week after irradiation, there was little time for annealing of potential damage. Therefore, we believe that our procedure represents a conservative testing approach.

\begin{figure}[htb]
	\begin{center}
		\includegraphics[height=4cm]{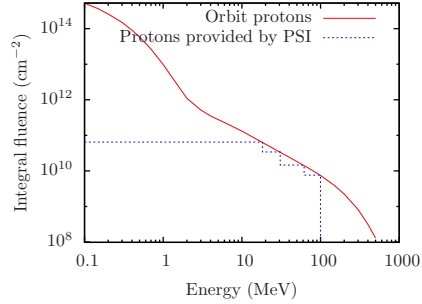}
	\end{center}
	\caption{Blue line: ``unit'' fluence provided by the PSI irradiation (after the degraders, sum of all irradiation steps, before the 10 mm Al shield). Red line: integral proton fluence in space on the STE-QUEST orbit.}
	\label{fig:radA}
\end{figure}

Considering that protons with energy lower than 20 MeV would be effectively stopped by a 2 mm aluminum shield, the mirrors were irradiated with protons with energies on the order of and higher than 20 MeV. The mirrors were irradiated in sequence with 4 energies, 18.29, 30.74, 61.6, and 99.7 MeV with corresponding ``unit'' fluences of $3.05\times10^{10}$, $1.96\times10^{10}$, $7\times10^{9}$, and $7.61\times10^{9}$ protons/${\rm cm}^{2}$. The energy 99.7 MeV corresponds to the input proton beam energy as delivered from the accelerator at PSI, while the first three values are the mean energies after degradation by copper plates (``degraders'') of thickness 12.5, 11.5, and 7.5 mm, respectively. The integral fluence of the proton irradiation agrees with the integral fluence of the STE-QUEST mission in the energy range 15 - 100 MeV. The energy range above 100 MeV could not be implemented, due to the energy limitation of the proton accelerator. Figure \ref{fig:radA} shows the PSI-produced integral fluence as a blue line (sum over all irradiation steps).

In this irradiation, we simulated the assumed existence of 10 mm thick aluminum shielding in the spacecraft by inserting an additional 10 mm thick aluminum plate in front of the mirrors.

\begin{figure}[htb]
	\centering
	\includegraphics[height=4cm]{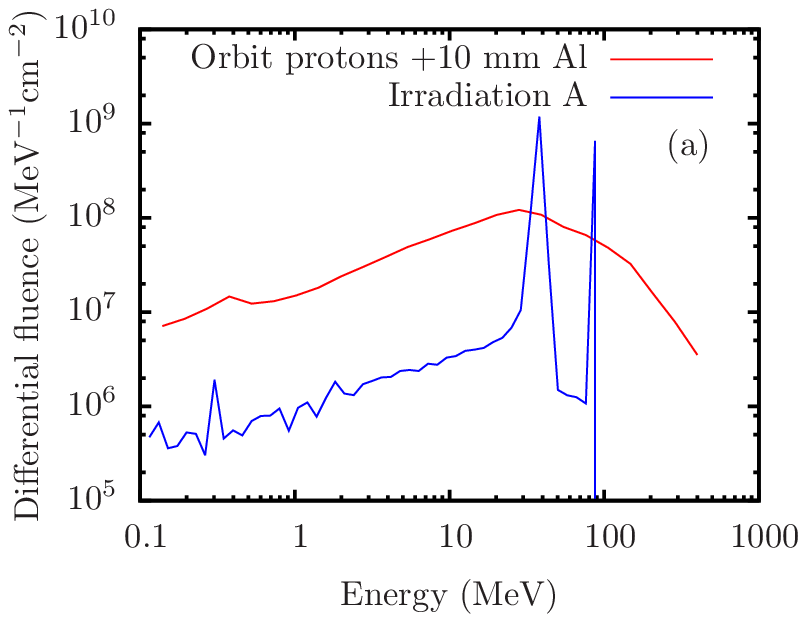}~~\includegraphics[height=4cm]{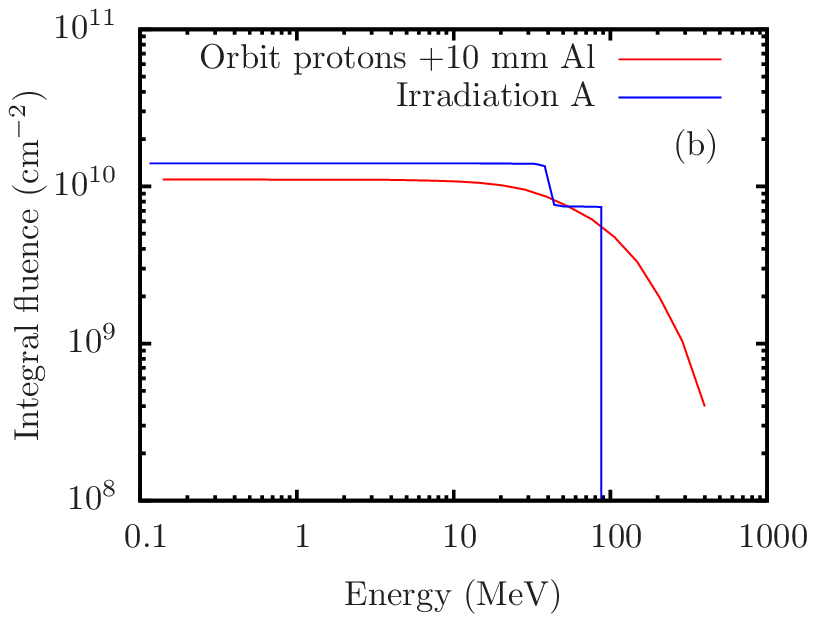}
	\caption{(a) Red line: simulated STE-QUEST differential proton spectrum after 10 mm aluminum shielding. Blue line: simulated spectrum of the ``unit'' proton irradiation delivered at PSI after the 10 mm aluminum filter (all irradiation steps combined). (b) same as (a), but showing the energy-integrated fluences.}
	\label{fig:radAshield}
\end{figure}

The spectrum of the total proton irradiation after the 10 mm Al shielding is calculated by using MULASSIS (60 energy bins were used in the calculation). The result is shown in Fig. \ref{fig:radAshield} (Irradiation A). The 99.7 MeV input energy is reduced to approximately 90 MeV, while the 61.6 MeV mean energy after the copper retarder is reduced to 40 MeV mean energy. The figure also shows that there are no peaks near 18 and 31 MeV; the protons produced in these two irradiation steps are stopped by the shielding. Note that the integral proton fluence (Fig. \ref{fig:radAshield} (b), blue line) agrees with the fluence of STE-QUEST mission. However, the differential fluence of the proton irradiation at low energy ($<$ 10 MeV) (Fig. \ref{fig:radAshield} (a), blue line) is much lower than the STE-QUEST spectrum. A fraction of the low-energy protons that will be present in space (red line in Fig. \ref{fig:radAshield} (a)) may be stopped and incorporated inside the mirrors and therefore may have an important effect on the mirrors, such as volume swelling. Therefore, a second proton irradiation was carried out, aimed at producing a suitable low-energy proton spectrum.

\begin{figure}[htb]
	\centering
	\includegraphics[height=4cm]{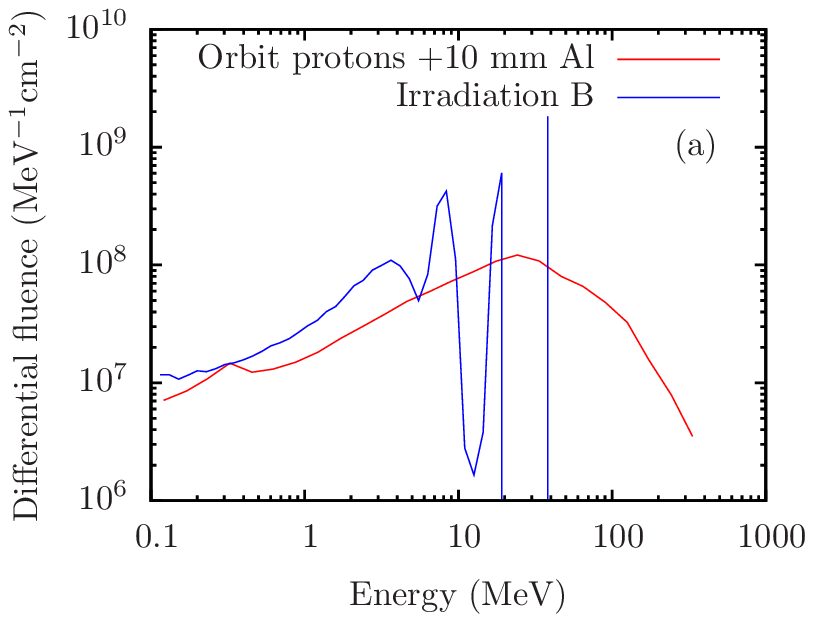}~~\includegraphics[height=4cm]{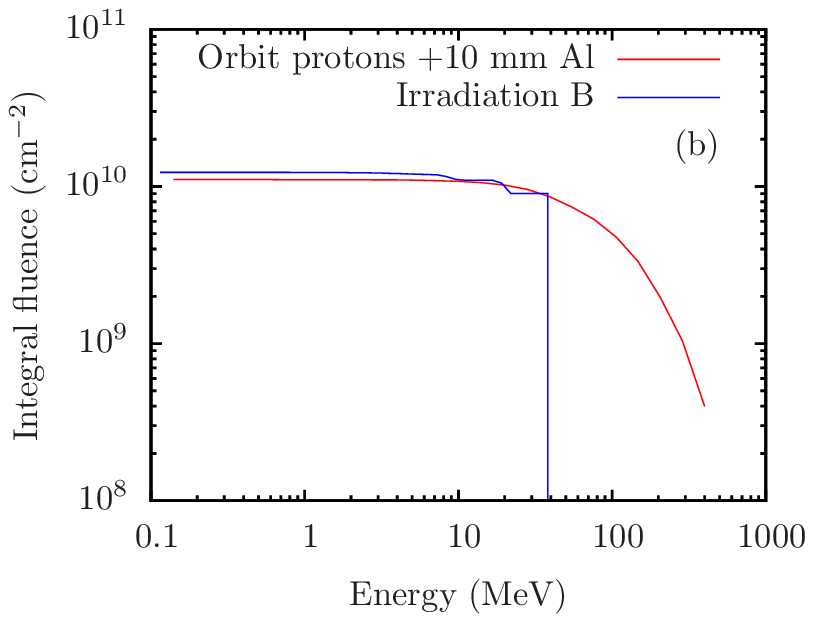}
	\caption{Simulated fluences (blue lines) for the ``unit'' irradiation carried out at the University of Birmingham.}
	\label{fig:radB}
\end{figure}

The second, complementary proton irradiation was carried out in the University of Birmingham (UB). The maximum energy of the proton beam in the accelerator is 36 MeV. We designed aluminum degraders to reproduce the STE-QUEST proton irradiation spectrum at low energy. With aluminum degrader thicknesses of 0, 4.5, 5.7, and 6 mm placed in the beam, and corresponding ``unit'' irradiation fluences of $9\times10^{9}$, $2\times10^{9}$, $1\times10^{9}$, and $4\times10^{8}$ protons/cm$^{2}$, the differential and integral proton fluences are shown in Fig. \ref{fig:radB} (Irradiation B), obtained using MULASSIS simulations. Now, both the differential and integral proton fluences are in reasonable agreement with the STE-QUEST proton spectrum after a 10 mm aluminum shield. We emphasize that in the UB irradiation, there was no additional 10 mm shield placed before the mirrors.

Three mirrors, which were marked with 1, 2, and 3 for identification, were irradiated with 1/5, 3/5, and 5/5 of the unit fluence ``A'' at PSI, by interrupting the irradiation twice and removing first mirror 1 and later mirror 2, which correspond to 1, 3, and 5 years permanence in the STE-QUEST orbit  behind a 10 mm Al shield.

\begin{figure}[tbh]
	\centering
	\includegraphics[height=4cm]{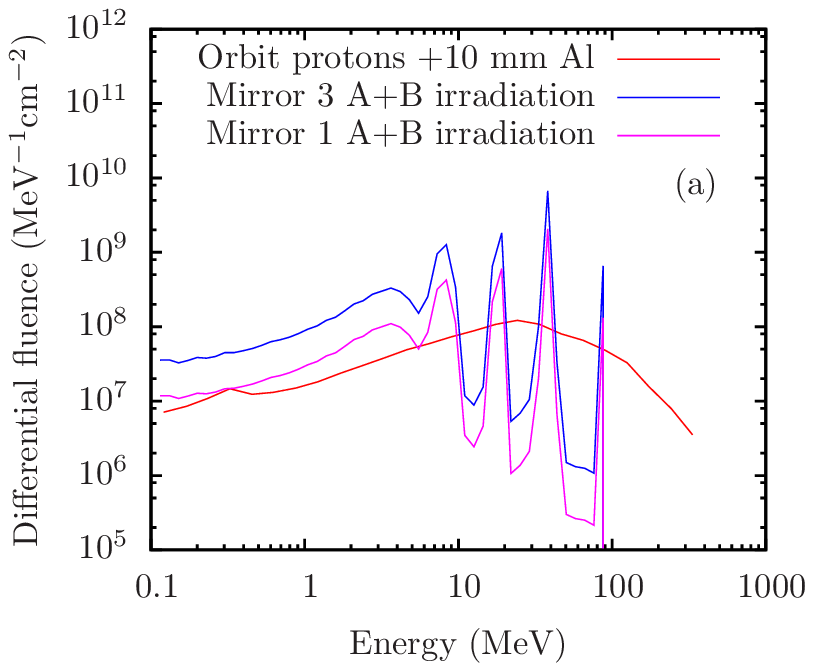}
	\includegraphics[height=4cm]{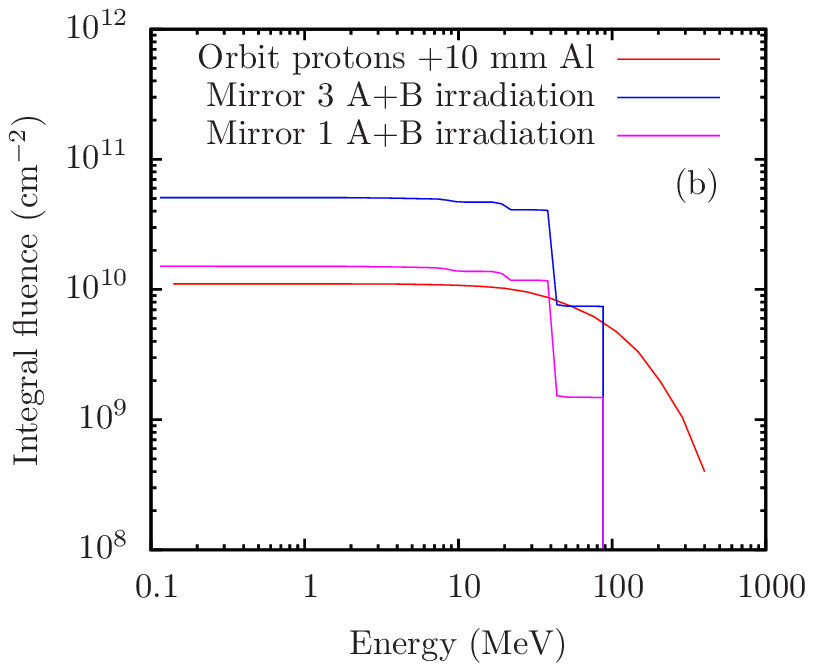}
	\caption{The total proton irradiation applied on the test mirrors. The irradiation applied on mirror 1 was 1/5 unit of irradiation A plus 1 unit of irradiation B. The irradiation applied on mirror 3 was 1 unit of irradiation A plus 3 units of irradiation B. }
	\label{fig:radSum}
\end{figure}

5 months later, the three mirrors were irradiated at UB. The mirrors were irradiated with 1, 2, and 3 times the unit fluence ``B'' at UB, which corresponds to 5, 10, and 15 years permanence in the STE-QUEST orbit behind a 10 mm shield, respectively. The total irradiations applied on the test mirrors are shown in Fig. \ref{fig:radSum}.

{The mirrors were usually stored in the plastic box in which the mirrors were delivered from the coating manufacturer. During shipment and irradiation, instead, the mirrors were fixed to half-inch lens mounts and covered with thin plastic foils and lens tissue to protect their HR coating.}

\section{Optical setup for mirror characterization}

\begin{figure}[htb]
	\centering
	\includegraphics[width=8cm]{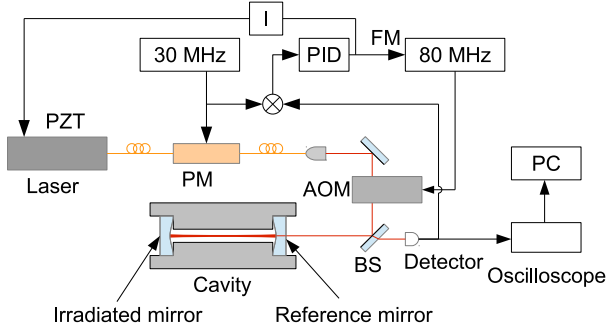}
	\caption{Optical setup for the determination of the cavity properties. BS: beam splitter, PM: waveguide phase modulator, AOM: acousto-optic modulator {(ISOMET 1205C)}, PZT: piezoelectric frequency tuning element, FM: frequency modulation input to the synthesizer {(Agilent E4420B)}, PID: proportional-integral-derivative circuit, I: integral circuit.}
	\label{fig:setup}
\end{figure}

The optical setup used to measure the linewidth and transmission of a cavity is shown in Fig. \ref{fig:setup}. The cavity was built by fixing two mirrors to the ends of a 50 mm long aluminum tube. It was not evacuated. A fiber laser at 1.5 $\mu$m was coupled to the fundamental mode of the cavity. A waveguide phase modulator (PM) generated the optical sidebands for the Pound-Drever-Hall frequency locking technique. An acousto-optic frequency shifter (AOM) was used to implement a fast feedback for the frequency stabilization of the laser and also suppressed the optical interference between the laser setup and the cavity. 

\begin{figure}[htb]
	\centering
	\includegraphics[height=4cm]{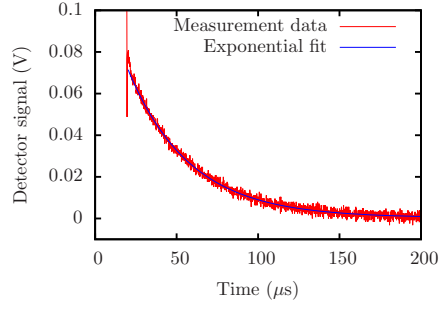}
	\caption{An example ring-down signal recorded by the reflection detector obtained by switching the input laser wave off while the laser frequency is in lock. The decay time of the cavity obtained by exponential decay fitting of this data is 37.6 $\pm$ 0.2 $\mu$s, equivalent to a finesse 708 000. The spike at the very beginning of the decay trace is caused by the strong reduction of the laser power reflected from the cavity mirror while the laser frequency is in resonance with the cavity. The imperfect fit at the beginning of the decay trace is caused by the limited RF switching speed of the AOM driver.}
	\label{fig:decay}
\end{figure}

The linewidths of the cavities were determined by measuring the decay times. To this end, the laser was switched off by switching off the RF output of the AOM driver while the laser frequency was locked to the resonance of the cavity. The detector in the optical setup, which normally recorded the reflection from the cavity, then recorded the decay of the laser power leaking out of the cavity. Figure \ref{fig:decay} shows an example of the signal measured in this way. 

The signal at the starting point of the cavity decay time trace gives the transmission out of the cavity, if the optical properties of the mirrors are the same. The coupling efficiency ($C$) and throughput ($T$) of the cavity are calculated from the laser power reflected from the cavity with the laser frequency in lock ($I_{r}$) and unlocked (off-resonance) ($I_{i}$) and the transmission of the cavity ($I_{t}$) by using the following formulae:

\begin{eqnarray*}
	C&=&{4 (I_{i}-I_{r})I_{t}\over(I_{i}-I_{r}+I_{t})^{2}}\,;\\
	T&=&{4I_{t}^{2}\over(I_{i}-I_{r}+I_{t})^{2}}\,.
\end{eqnarray*}

\section{Procedure}
The procedure to test the influence of the irradiation on the high-finesse mirrors was as follows. Three sample mirrors were irradiated. One mirror, which was not irradiated, was used as the reference mirror. The reference mirror combined with the irradiated mirrors 1, 2, 3 was used to build three high-finesse Fabry-P\'erot cavities. The linewidths of the cavities were measured by using the ring-down method before and after proton irradiation. Before assembling the cavities and measuring their linewidths, the mirrors were carefully cleaned by using lens tissue with spectroscopy-grade ethanol and observation of the mirror surface under a microscope.

\section{Results and discussion}
The linewidths of the cavities built from the mirrors before irradiation are about 4.3 kHz (Table 1, second column). The corresponding finesses of the cavities are about 700 000.

The linewidths of the cavities built by the reference mirror and the three mirrors after irradiation at PSI are shown in Table 1, third column. The linewidths of the cavities after the additional irradiation at UB are shown in the last column.

\begin{table}
	\centering
	\caption{Linewidths of the optical cavities before and after the proton irradiation. The cavities were 5 cm long and contained one unirradiated reference mirror and one of the three listed irradiated mirrors.}
	\begin{tabular}{c|p{3cm}|p{3cm}|p{3cm}}
		\hline
		Mirror No. & Linewidth \par before irradiation & Linewidth after \par irradiation at PSI & Linewidth after\par irradiation at UB \\
\hline
1 &
4.3 kHz &
4.3 kHz \par (1/5 $\times$ fluence A) &
4.4 kHz \par (1 $\times$ fluence B) \\
\hline
2 &
4.3 kHz &
4.5 kHz \par (3/5 $\times$ fluence A) &
4.5 kHz \par (2 $\times$ fluence B) \\
\hline
3 &
4.2 kHz &
4.4 kHz \par (5/5 $\times$ fluence A) &
4.3 kHz \par (3 $\times$ fluence B) \\
\hline
	\end{tabular}
	\label{tab:result}
\end{table}
The table shows that before and after the maximum irradiation fluence (mirror 3) the change of the linewidth of the cavity is less than 100 Hz (single irradiated mirror). If both mirrors were irradiated, the change would be less than 200 Hz. This is less than 5\% in relative terms. It is within the {uncertainty of the cleaning procedure of the mirror surfaces. The repeatability of the cleaning procedure was tested before the irradiation test was carried out. We found that with the same pair of mirror the linewidth varies between 4.2 kHz and 4.4 kHz after each cleaning. The coupling efficiencies before and after the irradiation are all about $C = 70$\% and the throughputs of the cavities are about $T = 20$\%. 

{Concerning the proton irradiation test, there are three main differences between our test and a future operation of the resonator in space. First, the proton energy spectrum in our experiment is discrete at high energy and reaches only up to about 90 MeV, while the spectrum in space extends beyond 500 MeV. Although the experimental spectrum did not cover the full spectrum, we believe that protons with energy higher than 90 MeV may not have a stronger effect on the mirrors than 90 MeV protons, because this latter energy is already ten times higher than the binding energy of a nucleon in the nucleus, and because the mirrors (or coatings) are not thick enough to stop the protons. Second, the irradiation time (approx. 2 days) in the experiment is much shorter than the duration of a space mission (years). We believe that, at the same fluence, the shorter irradiation time would cause stronger degradation, if any, because much less time is given to the mirror coatings to anneal damage. Third, in our experiment, the mirrors were stored and irradiated in air, while in a space application, the resonator will operate in a high-vacuum chamber. The mirror coatings may undergo reactions with oxygen or moisture in air, or be scratched during the cleaning procedure in the experiment, effects that are excluded when a resonator operates in a vacuum chamber. Therefore, we believe that the downgrade of the mirror coating caused by proton radiation when the resonator is operated in the space on a STE-QUEST-like orbit and duration, will not be underestimated in this test.}

\section{Conclusion}

This investigation showed that the effect of 15 years STE-QUEST-equivalent low-energy proton irradiation plus 5 years high-energy irradiation is less than 5\% on a cavity with finesse of 700 000 in the near-infrared, and is also not observable on the coupling efficiency and throughput of the cavity.  This result is based on the presence of a 10 mm equivalent Al shielding around the mirror, a realistic assumption. This represents an encouraging result for advanced space missions requiring high-performance frequency metrology.

Future investigations should address the sensitivity of the coatings {to high energy protons ($>$90 MeV) and} to electron irradiation. Furthermore, high-finesse cavities are often constructed with a spacer made from an ultra-low-expansion material. The effect of proton irradiation on the temperature of zero thermal expansion{,} and on the long-term drift of the length of such a cavity is an important open question, motivated by the observation of property changes upon electron irradiation. {Finally, an important further issue to be investigated is the possible change of the mechanical loss (damping) of the mirror substrate and of the coating. This may lead to a change of the thermal noise level of the cavity resonance frequencies \cite{Numata:2004}. This is a potential issue for optical cavities for the most sophisticated space applications, such as optical clocks.}

\section*{Acknowledgment}

	We thank Advanced Thin Films (USA) for partial support of this work. We thank D. Iwaschko and P. Dutkiewicz for electronics development. We are indebted to M. Gehler and U. Sterr for discussions, and to K. Bongs for support. The research leading to these results has received funding from the Bundesministerium f\"ur Wirtschaft und Technologie (Germany) under project no. 50OY1201. The CSEM's authors gratefully thank the Swiss Space Office, the Prodex office and the Canton of Neuch\^{a}tel for financial support. 


\end{document}